\documentclass[conference]{IEEEtran}
\IEEEoverridecommandlockouts
\usepackage{cite}
\usepackage{amsmath,amssymb,amsfonts}
\usepackage{algorithmic}
\usepackage{graphicx}
\usepackage{textcomp}
\usepackage{xcolor}
\usepackage{subcaption}
\usepackage{tabularx, booktabs}
\usepackage[braket,qm]{qcircuit}
\usepackage{algorithmic}
\usepackage[ruled, linesnumbered]{algorithm2e}
\usepackage{multirow}
\usepackage{balance}
\SetKwComment{Comment}{\# }{}

\def\BibTeX{{\rm B\kern-.05em{\sc i\kern-.025em b}\kern-.08em
    T\kern-.1667em\lower.7ex\hbox{E}\kern-.125emX}}
\begin{document}

\bstctlcite{IEEEexample:BSTcontrol}

\title{Quantum Masked Autoencoders for Vision Learning}

\author{\IEEEauthorblockN{Emma Andrews and Prabhat Mishra}
\IEEEauthorblockA{
\textit{University of Florida, 
Gainesville, FL, USA}}
}

\maketitle

\begin{abstract}
Classical autoencoders are widely used to learn features of input data. To improve the feature learning, classical masked autoencoders extend classical autoencoders to learn the features of the original input sample in the presence of masked-out data. While quantum autoencoders exist, there is no design and implementation of quantum masked autoencoders that can leverage the benefits of quantum computing and quantum autoencoders. In this paper, we propose quantum masked autoencoders (QMAEs) that can effectively learn missing features of a data sample within quantum states instead of classical embeddings. We showcase that our QMAE architecture can learn the masked features of an image and can reconstruct the masked input image with improved visual fidelity in MNIST-family images. Experimental evaluation highlights that QMAE can significantly outperform (12.86\% on average) in classification accuracy compared to state-of-the-art quantum autoencoders in the presence of masks.
\end{abstract}


\section{Introduction} \label{sec:intro}
Feature learning is a critical task to understand the data samples during machine learning. This enables the model to establish relationships between the features within the data samples and aid the model in its specific task, such as classification. Autoencoders are a popular model architecture for learning and extracting the features within data, taking the original data and compressing it into a smaller data space~\cite{hinton2006reducing, bank2021autoencoders}. If the autoencoder learns the features, it can then take the compressed data and reconstruct it back to the original data space, with minimal loss of information.

In classical machine learning, there are concerns with the scalability of models and data, meaning that as the models grow in parameters and the data grows in feature size, they become computationally expensive to train and run. Autoencoders are no exception, becoming increasingly costly to process the input data and handle the reconstruction. Thus, ensuring that the model can learn the features under the presence of less data is crucial. 


Masked autoencoders (MAEs) are a variant of autoencoders that learn features efficiently, even under the presence of masked, or missing, information from the original data sample~\cite{he2022masked}. These are able to achieve feature learning despite missing 70-80\% of the original data sample, such as an image. The masked autoencoder can successfully fill in masked information in the original data sample without the data originally being there during reconstruction.

Quantum machine learning (QML) can produce machine learning models that can achieve comparable results to classical machine learning models using fewer parameters. Quantum autoencoders (QAEs) can create models that can compress and learn data samples in a fraction of the parameters compared to their classical counterparts~\cite{romero2017quantum}. For traditional QAEs, masking out information in the original data sample results in the mask being present after the reconstruction, not learning the features in the masked regions. 

In this paper, we propose quantum masked autoencoders (QMAEs) to efficiently learn the features and reconstruct the entire original input, even under the presence of masked regions. Specifically, this paper makes the following contributions.
\begin{itemize}
    \item We develop a QMAE architecture, learning features of masked information present in image samples in the MNIST~\cite{lecun1998mnist}, FashionMNIST~\cite{xiao2017fashionmnist}, and Kuzushiji-MNIST~\cite{clanuwat2018deep} datasets. To the best of our knowledge, this is the first work establishing masked autoencoders as a quantum machine learning model.
    The original images can be masked up to 25\% and the reconstruction can achieve good similarity with the original image. This similarity is also better than an equivalent QAE.
    \item We showcase that with the reconstructions from QMAE, they can be used in classifiers and have a greater accuracy (12.86\%) in label predictions compared to QAEs.
\end{itemize}

The rest of the paper is organized as follows. Section~\ref{sec:bg} surveys related work. Section~\ref{sec:qmae} presents the QMAE architecture, including training considerations such as the loss function. Section~\ref{sec:results} presents the experimental results. Finally, Section~\ref{sec:conc} concludes the paper.
\section{Background and Related Work} \label{sec:bg}
In this section, we describe relevant background and survey related efforts in both classical and quantum domains.

\subsection{Classical Autoencoders}
\begin{figure}[htp]
    \centering
        \vspace{-0.1in}
    \includegraphics[width=0.9\linewidth]{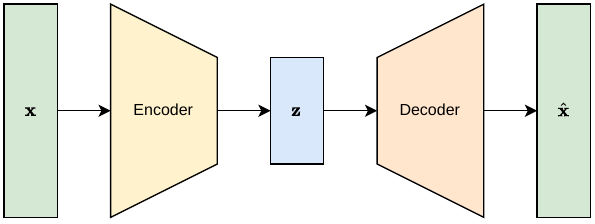}
    \caption{Classical autoencoder architecture.}
        \vspace{-0.1in}
    \label{fig:ae}
\end{figure}

Autoencoders were initially proposed in the classical domain, featuring an encoder and decoder~\cite{hinton2006reducing}. The encoder is responsible for processing the original input data $x$ into a smaller data space, also known as the latent space. The compressed result $z$ of the original input $x$ from the encoder can be given as input to the decoder to reconstruct back to the original data space $x$ exists in. The result from the decoder is $\hat{x}$, and is ideally as close to the original data sample $x$ as possible. This architecture is shown in Figure~\ref{fig:ae}.

Masked autoencoders (MAEs)~\cite{he2022masked} are a specific type of autoencoder that masks out or removes portions of the input data before encoding the data. During reconstruction, this missing data is reconstructed from the knowledge of surrounding information and through a learnable mask token. This learnable mask token is inserted into the logical patch locations of the masked patches in the latent space. Along with the encoded visible patches, the learnable mask token is given as input to the decoder to reconstruct the original data sample.


\subsection{Quantum Machine Learning}
Quantum computing operates on data that can be both 0 and 1 simultaneously, whereas classical computing data is either 0 or 1. The quantum state of a qubit can be expressed in superposition as
\begin{equation}
    |\psi\rangle=\alpha|0\rangle+\beta|1\rangle,
\end{equation}
where $\alpha$ and $\beta$ are the amplitudes satisfying $|\alpha|^2+|\beta|^2=1$. In a quantum circuit, the gates process this data by applying a unitary matrix onto the current quantum state value. A collection of these gates are parameterized gates, where the exact value of the unitary matrix is determined by the evaluation of the parameter. Quantum circuits that operate primarily with parameterized gates are known as variational quantum circuits (VQCs).

Entanglement is another important property in quantum computing where two or more quantum states are computed together such that they establish a relationship between them. This relationship can be leveraged across different applications, such as machine learning.

QML trains models on datasets to carry out a specific task, such as classification~\cite{schuld2015introduction, biamonte2017quantum}. It constructs the models through VQCs, serving as the ansatz. The parameters in the gates of the VQCs are taken as the parameters of the model, which are optimized during the training process towards an objective via a loss function. This training process is often done classically, using the same techniques to optimize and train classical machine learning models.

\subsection{Related Work}
\begin{figure}[h]
    \centering
    \includegraphics[width=\linewidth]{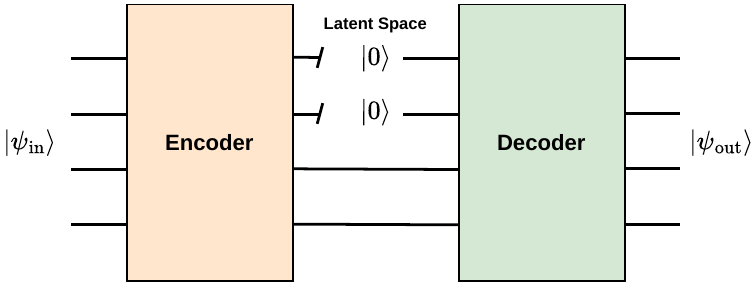}
    \vspace{-0.1in}
    \caption{Quantum autoencoder architecture.}
    \label{fig:qae}
\end{figure}

Similar to classical autoencoders, quantum autoencoders (QAE)~\cite{romero2017quantum} utilize an encoder and decoder. Both are implemented as variational quantum circuits, encoding the input quantum state data into the latent space then decoding it back to a reconstruction of the original data sample. Figure~\ref{fig:qae} provides an overview of this architecture.

For example, QAEs can compress and reconstruct images to a high quality~\cite{wang2024quantum}. This is achieved with a specific ansatz for the encoder and the decoder that lends itself to entangling the qubit states. However, if masked data were to be given as input to a QAE with the goal of reconstructing the masked image, the mask would be reconstructed as that missing information is seen as a feature of the original image. 

Note that it is not possible to directly use the concepts of classical MAEs to develop quantum MAEs due to intricacies with quantum circuits. This is primarily due to limitations with state preparation and measurements during the middle of an executing quantum circuit. Therefore, we need to adapt the classical MAE architecture to quantum to have similar resulting models in the quantum domain.


There are many applications of QAEs and their variants that perform feature learning and extraction. This includes anomaly detection~\cite{madeira2024quantum, tscharke2025quantum}, quantum error correction~\cite{locher2023quantum}, and quantum circuit processing~\cite{wu2024quantum}. However, none of these applications explore the presence of masked information and learning features to reconstruct information that is masked. To the best of our knowledge, there are no prior efforts in developing quantum autoencoders that can deal with data in the presence of masks.
\section{Quantum Masked Autoencoder (QMAE)} \label{sec:qmae}
We propose a quantum masked autoencoder (QMAE) that can reconstruct original data in the presence of masks. Figure~\ref{fig:arch} shows the QMAE architecture that consists of four major components to learn the masked information: image embedding (Section~\ref{sec:embed}), the encoder and decoder ansatz (Section~\ref{sec:ansatz}), the learnable mask token (Section~\ref{sec:token}), and the training including the loss function (Section~\ref{sec:loss}). The remainder of this section details these four components.

\subsection{Image Embedding} \label{sec:embed}
Images consist of pixel values which are classical in nature. Thus, these values must be embedded into quantum states so that quantum circuits can process the data. We utilize amplitude embedding~\cite{schuld2018supervised}, flattening the image into a one-dimensional vector of size $2^n$ such that
\begin{equation}
    |\psi\rangle=\sum^{2^n}_{i=0}x_i|i\rangle,
\end{equation}
where $n$ is the number of qubits required and $x_i$ is the $i$th vector element in the one-dimensional vector $x$ of the image data. It is important to note that the amplitude embedding will normalize the data if it is not already normalized.

\subsection{Encoder and Decoder Ansatz} \label{sec:ansatz}
The encoder and decoder ansatz carry out the main computation of the model and both variational quantum circuits, parameterized by $\theta$. Thus, the encoder is defined as the ansatz $U(\theta)$ and the decoder is its adjoint, $U^\dagger(\theta)$. Note that the decoder is also parameterized by the same values $\theta$ as the encoder. Figure~\ref{fig:qae} shows the interaction that the encoder and decoder ansatz have together. The encoder ansatz takes in the input state $|\psi_{\text{in}}\rangle$, consisting of $n$ qubits. The encoder compresses $|\psi_{\text{in}}\rangle$ into $k$ qubits, or the latent space, where $k<n$. The remaining $t=n-k$ qubits consist of the trash space and are discarded, or reset to $|0\rangle$, prior to running the decoder. Thus, the decoder reconstructs the compressed $|\psi_{\text{in}}\rangle$ into $n$ qubits and results in the state $|\psi_{\text{out}}\rangle$.

\begin{figure}[h]
    \centering
    \includegraphics[width=\linewidth]{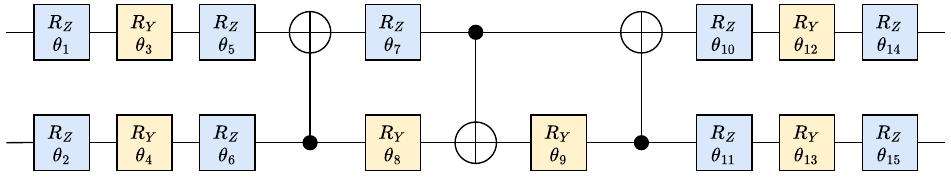}
    \caption{Two-qubit interaction circuit, originally proposed for image compression QAEs by Wang et al.~\cite{wang2024quantum}. The circuit consists of 9 parameterized $R_Z$ gates, 6 parameterized $R_Y$ gates, and 3 CNOT gates.}
    \label{fig:twolocal}
\end{figure}

We utilize the ansatz presented by Wang et al.~\cite{wang2024quantum} as the encoder and decoder ansatz to strongly entangle pairs of qubits together. This involves a two-qubit interaction circuit consisting of 18 gates with 15 parameters. The rotation gates are parameterized and consist of 3 CNOT gates to entangle the result together. Therefore, the parameters are minimized while the qubits are entangled together. This two-qubit interaction circuit is presented in Figure~\ref{fig:twolocal}. Each qubit pair within the data input range processes this two-qubit interaction circuit. Therefore, the total number of parameters that must be trained for a data input spanning $n$ qubits is 
\begin{equation}
    p_a=\frac{n(n-1)}{2}\times15.
\end{equation}

\begin{figure}
    \centering
    \includegraphics[width=\linewidth]{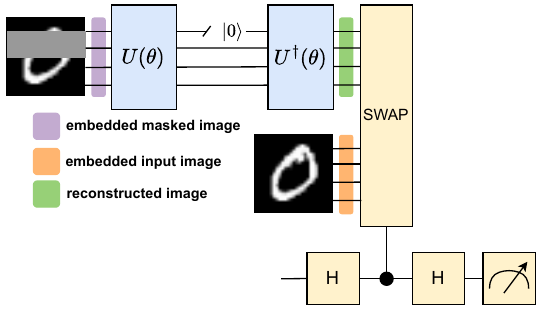}
    \caption{QMAE architecture. The original image is masked and embedded as input to the encoder $U(\theta)$. The decoder takes the compressed representation and reconstructs the image, with learned features. A SWAP test is then performed between the reconstructed image and the original image to get the fidelity.}
    \label{fig:arch}
\end{figure}

The overall architecture is showcased in Figure~\ref{fig:arch} for input data requiring 4 qubits to process, such as an $8\times8$ image. The architecture first embeds the masked image into a quantum state through an amplitude embedding, as described in Section~\ref{sec:embed}, onto qubits 2-5. 

We can denote the different qubits and their functionality as different subsystems. Subsystem $A$ is the latent space qubits, subsystem $B$ is the trash space qubits, and subsystem $B'$ is the reference space qubits. The reference space qubits contain a reference to the state that the output is compared against, such as for fidelity measurements. Thus, the output of the encoder ansatz can be defined as
\begin{equation}
    U(\theta)|\psi\rangle_{AB}=|\psi\rangle_A\otimes|\psi\rangle_B,
\end{equation}
where $|\psi\rangle_A$ is the compressed state on the latent space qubits, $|\psi\rangle_B$ is the trash state on the trash qubits, and $|\psi\rangle_{AB}$ is the complete output from the encoder, including both the latent and trash qubits.

Ideally, if the encoder learns the latent space representation of $|\psi\rangle_{AB}$, then the trash state $|\psi\rangle_B$ being reset to $|0\rangle$ results in the decoder outputting the original state $|\psi\rangle_{AB}$, or
\begin{equation}
    U^\dagger(\theta)U(\theta)|\psi\rangle_{AB}=|\psi\rangle_{AB}.
\end{equation}
However, as it must learn this representation, this results in the density matrix from the full encoder-decoder ansatz of
\begin{equation} \label{eq:qae} 
    \rho_{\text{out}}=U^\dagger_{AB'}\left(\theta\right)\text{Tr}_B\left[U_{AB}\left(\theta\right)|b\rangle U_{AB'}\left(\theta\right)\right]U_{AB'}\left(\theta\right),
\end{equation}
where $|\psi_{in}\rangle$ is the input state to the encoder or the $AB$ system, $|b\rangle=|\psi_{in}\rangle\langle\psi_{in}|_{AB}\otimes|a\rangle\langle a|_{B'}$, and $|a\rangle$ is the reference state contained in $B'$.

\subsection{Learnable Mask Token} \label{sec:token}
As detailed in the ansatz architecture, the encoder input is the image with masks in place. For QMAE, instead of masking the data to a zero value, a learnable mask token is used instead. This learnable mask token is a parameter of the model of the total patch size and attempts to learn an efficient representation of the input dataset to be able to reconstruct information in that logical patch correctly, according to the original image with no masks. Thus, along with the parameters for the ansatz, the total model will train
\begin{equation}
    p=p_a+H_pW_p,
\end{equation}
where $H_p$ is the height of the patch and $W_p$ the width of the patch when the original image is considered as a two-dimensional image.

Insertion of the learnable mask token during input to the encoder allows for efficient circuit running, allowing the masked patches to be traded out for the learnable mask token prior to embedding the input to the circuit, akin to the MAE architecture presented in~\cite{chen2025masked}. Thus, the mask token represents classical space, matching the original input modality. Additionally, this causes a reduction in mid-circuit measurement and state preparation that would have to occur in between the encoder and decoder if the learnable mask token were to be inserted in the latent space, as is done in the original MAE architecture~\cite{he2022masked}.

\subsection{Training and Loss Function} \label{sec:loss}
\begin{figure}[h]
    \centering
    \includegraphics[width=0.7\linewidth]{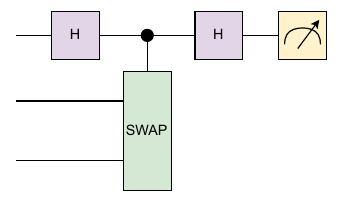}
    \caption{SWAP test to measure fidelity of two states located on qubit wires 1 and 2. }
    \label{fig:swap}
\end{figure}

To train QMAE, including updates and optimization to the learnable mask token, the loss function needs to represent how well the model is at producing results that are equivalent to the original data sample, without masked patches. One way to measure the similarity between two quantum states is the SWAP test~\cite{aimeur2006machine, schuld2015introduction}. The SWAP test, shown in Figure~\ref{fig:swap}, measures the fidelity between two states $|\psi\rangle$ and $|\phi\rangle$~\cite{buhrman2001quantum}. An ancilla qubit is used to contain the result of the fidelity measurement, with the expectation value measured at the end, resulting in
\begin{equation}
    \langle\sigma_Z\rangle=|\langle\phi|\psi\rangle|^2.
\end{equation}
Thus, the resulting expectation value $\langle\sigma_Z\rangle$ contains how similar states $|\psi\rangle$ and $|\phi\rangle$ are and can be used as part of the loss function to optimize the circuit. 

Specifically, in QMAE, the SWAP test measures the fidelity between the original input image with no masks and the reconstruction resulting from the output of the decoder. As such, since the original input image is not embedded into the circuit as the input to the encoder, extra qubits are required to perform an additional embedding operation to embed the original input image into the circuit for the SWAP test. In Figure~\ref{fig:arch}, these extra qubits can be seen by embedding the original image into qubits 6-9. With the fidelity measured as the expectation value $\langle\sigma_Z\rangle$, it can be used in a loss function to guide the optimization of circuit parameters. Thus, the resulting loss function for training QMAE is
\begin{equation} \label{eq:loss}
    \mathcal{L}=1-\langle\sigma_Z\rangle,
\end{equation}
which is minimized. The loss can be optimized through classical optimization techniques, such as Adam~\cite{kingma2014adam}.

The training process for QMAE is presented in Algorithm~\ref{alg:qmae}. For each epoch and each sample in the training dataset $\mathcal{D}$, the masked image $m$ is prepared, including embedding it into the ansatz $Q(\theta)$ with the original image $d$. The ansatz $Q(\theta)$ is executed with this embedded masked image, resulting in the fidelity $\langle\sigma_Z\rangle$ between the reconstructed image and the original image. This is minimized in the loss function $\mathcal{L}$ and used to optimize the parameters $\theta$.

\begin{algorithm}
    \caption{QMAE} \label{alg:qmae}
    \SetAlgoLined
    \KwData{$\mathcal{D}$: training dataset, $Q(\theta)$: ansatz, $E$: epochs}
    \KwResult{$\theta$: trained model parameters}
    \For{$e\in E$}{
    \For{$d\in\mathcal{D}$}{
        $i=\text{get\_mask\_idx}()$ \Comment*[r]{Random sampling}
        $m=\text{insert\_mask\_token}(d)$\;
        $\text{image\_embed}(m,d)$\;
        $\langle\sigma_Z\rangle=\text{execute}(Q(\theta))$\;
        $\mathcal{L}=1-\langle\sigma_Z\rangle$\;
        $\text{optimize}(\theta, \mathcal{L})$\;
    }
    }
\end{algorithm}

Our resulting QMAE architecture utilizes all four components presented to learn the features in images effectively. Specifically, the images are embedded from classical to quantum data through amplitude embeddings, where the encoder and decoder process the data with the learnable mask token inserted into mask patches. This model is trained with a fidelity loss function to ensure that the output quantum state from the model is similar to the original input image, without any masks. 
\section{Experiments}
\label{sec:results}
This section evaluates the effectiveness of QMAE using the MNIST-family datasets. We also compare the QMAE results to a QAE with the same ansatz. 

\subsection{Experimental Setup}
All experiments were run using Python v3.13.5, PennyLane v0.42.3~\cite{bergholm2022pennylane}, and PyTorch v2.8.0~\cite{ansel2024pytorch}. Additionally, torchmetrics v1.8.2 provided some additional metric implementations. We tested our QMAE implementation on MNIST~\cite{lecun1998mnist}, FashionMNIST (FMNIST)~\cite{xiao2017fashionmnist}, and Kuzushiji-MNIST (KMNIST)~\cite{clanuwat2018deep}. Each image in the datasets is grayscale and resized to $16\times16$, resulting in 8 qubits required to embed the images into quantum states. The latent space is chosen to be 7 qubits, leaving 1 qubit for the trash space. An additional 8 qubits are required to embed the original input for the fidelity measurement, and an additional ancilla qubit to measure the fidelity. The trash space must also be reset to $|0\rangle$ as input to the decoder, resulting in an extra qubit to prepare the trash state qubit via a SWAP. Thus, the resulting QMAE circuit for these datasets contains 18 qubits.

For comparison, we also train a QAE with the same ansatz, masked input (excluding the learnable mask token), and hyperparameters to showcase how the mask token can learn the missing features and represent them well during reconstruction. The QAE results produced during training are compared against the original input images in the loss function. We analyze the reconstructed images from both models on the visual fidelity of the images and classification accuracy, as outlined in the following subsections.

\subsection{Comparison of Reconstructed Images}
\begin{figure*}[!h]
    \centering
    \begin{subfigure}{0.325\linewidth}
        \centering
        \includegraphics[width=\linewidth]{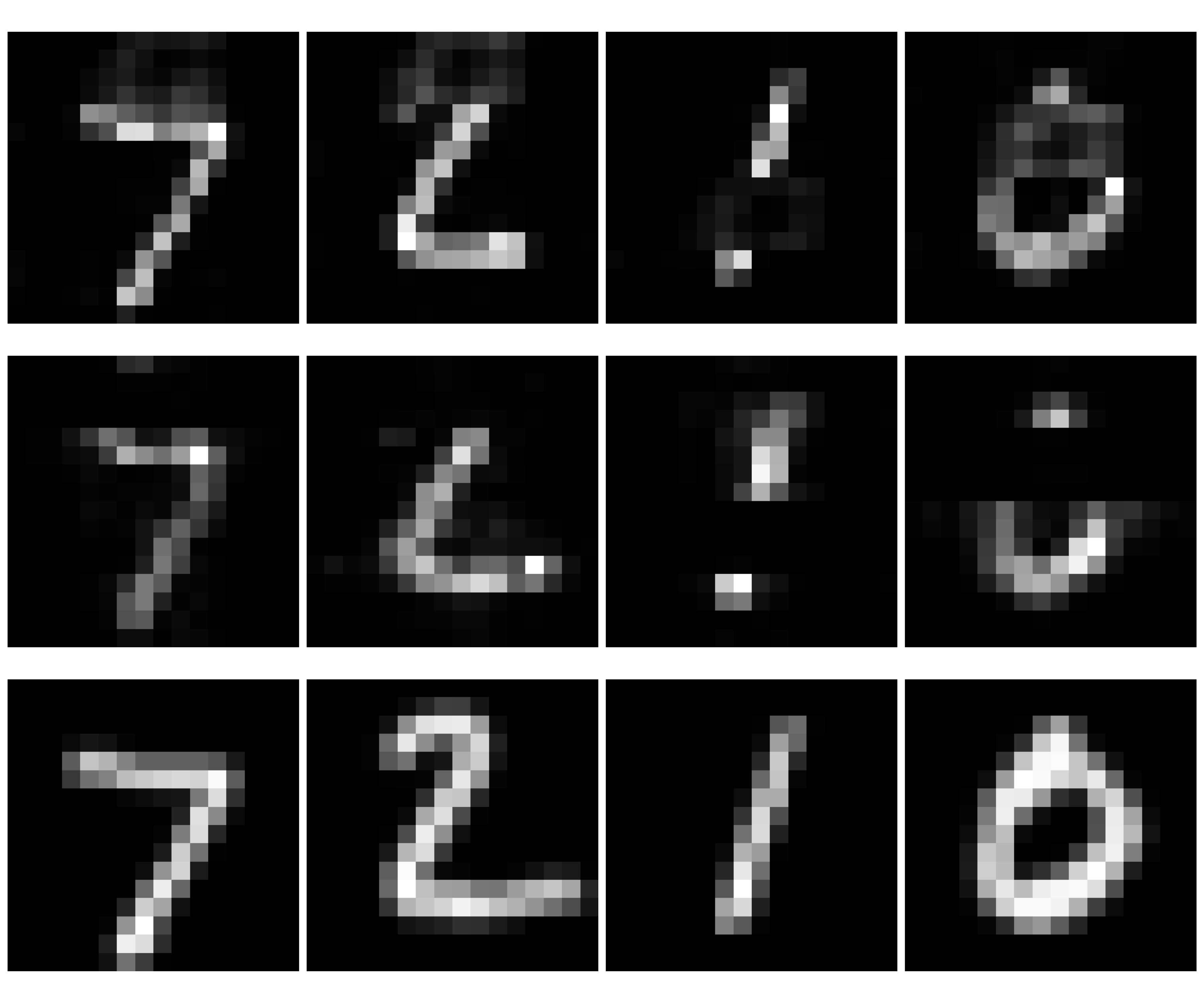}
        \caption{MNIST}
        \label{fig:mnistc1}
    \end{subfigure}
    \begin{subfigure}{0.325\linewidth}
        \centering
        \includegraphics[width=\linewidth]{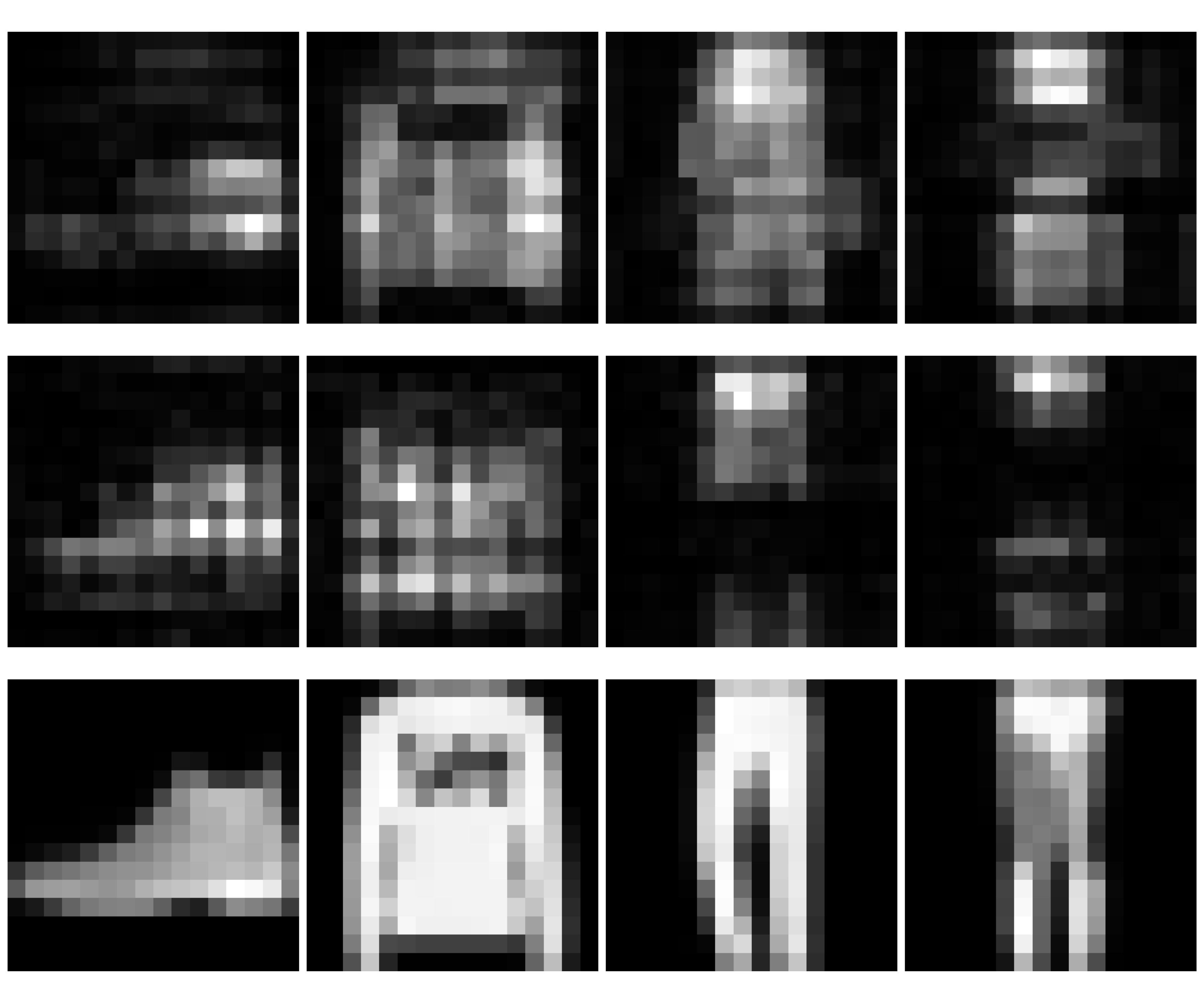}
        \caption{FMNIST}
        \label{fig:fmnistc1}
    \end{subfigure}
    \begin{subfigure}{0.325\linewidth}
        \centering
        \includegraphics[width=\linewidth]{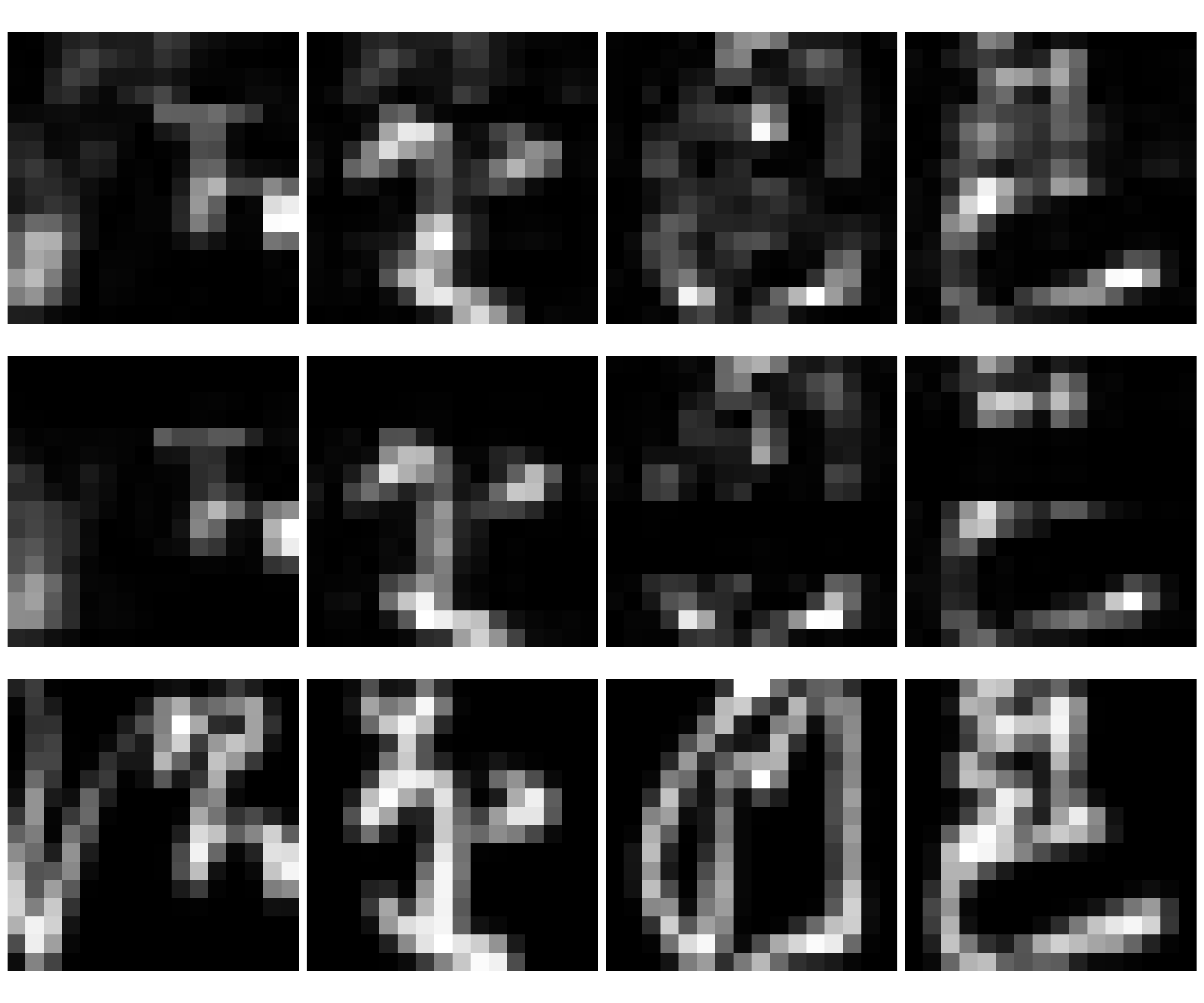}
        \caption{KMNIST}
        \label{fig:kmnistc1}
    \end{subfigure}
    \caption{Results from QMAE (top row) and QAE (middle row) at 25\% mask for the (a) MNIST, (b) FMNIST, and (c) KMNIST datasets. The original images are in the bottom row.}
    \label{fig:comp2}
\end{figure*}

Figure~\ref{fig:comp2} shows the original images (bottom row) as well as the reconstructed images (top two rows) by our QMAE architecture and the state-of-the-art QAE architecture under 25\% mask for the three datasets. 
Specifically, the bottom row (row 3) shows five original images (without masking) from each dataset. To enable a fair comparison between our proposed approach (QMAE) and state-of-the-art (QAE), we apply a single masked patch representing 25\% of the entire image. 
The results produced by QMAE with 25\% mask (row 1) are better in image quality than the results from QAE with 25\% mask (row 2) across all datasets. QAE is not able to learn any features in the masked areas, instead reconstructing the masked patch. In contrast, the QMAE model is able to learn the features to produce high-quality images in the presence of masked patches.


\begin{figure*}[!h]
    \centering
    \centering
    \begin{subfigure}{0.325\linewidth}
        \centering
        \includegraphics[width=\linewidth]{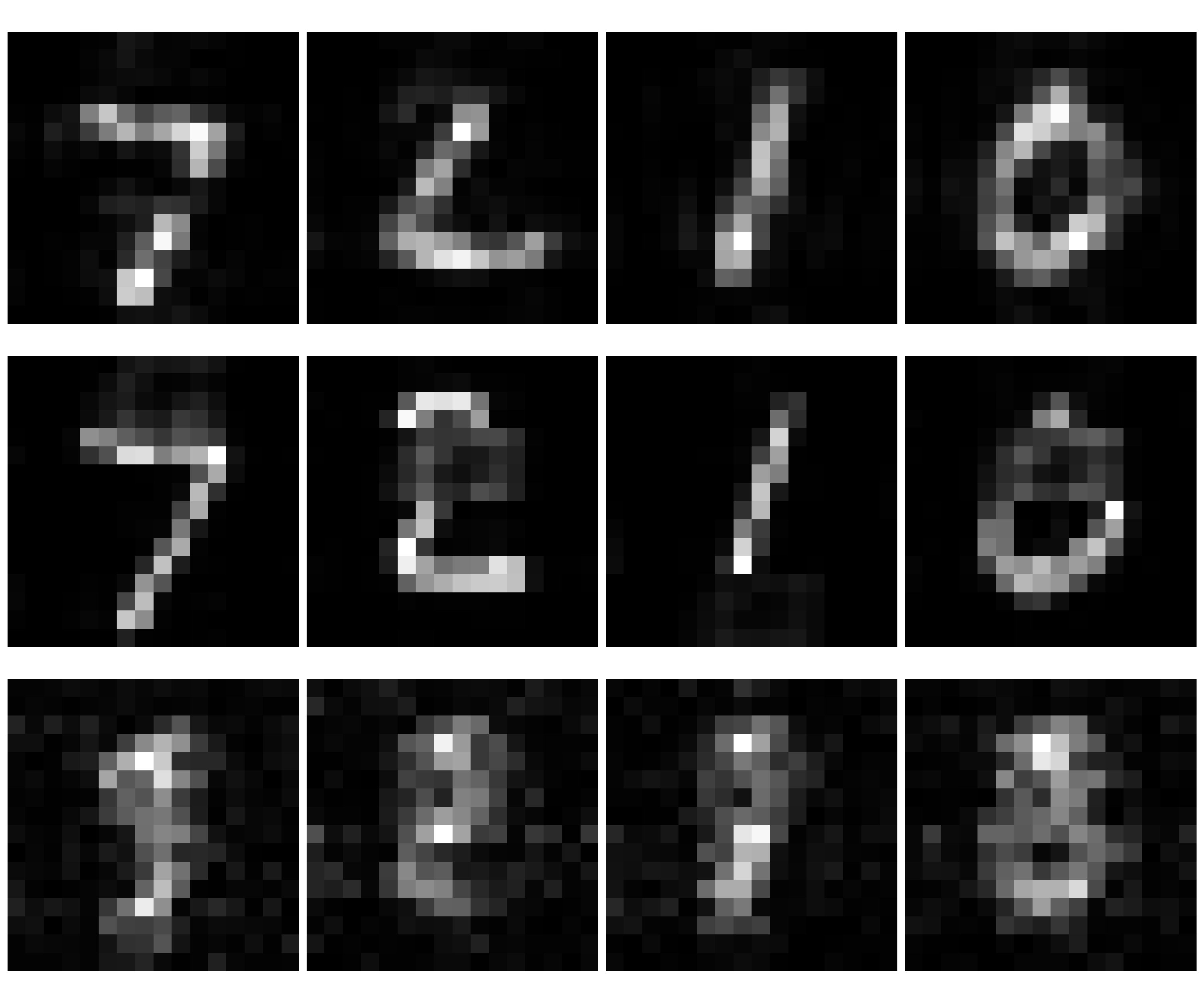}
        \caption{MNIST}
        \label{fig:mnistc2}
    \end{subfigure}
    \begin{subfigure}{0.325\linewidth}
        \centering
        \includegraphics[width=\linewidth]{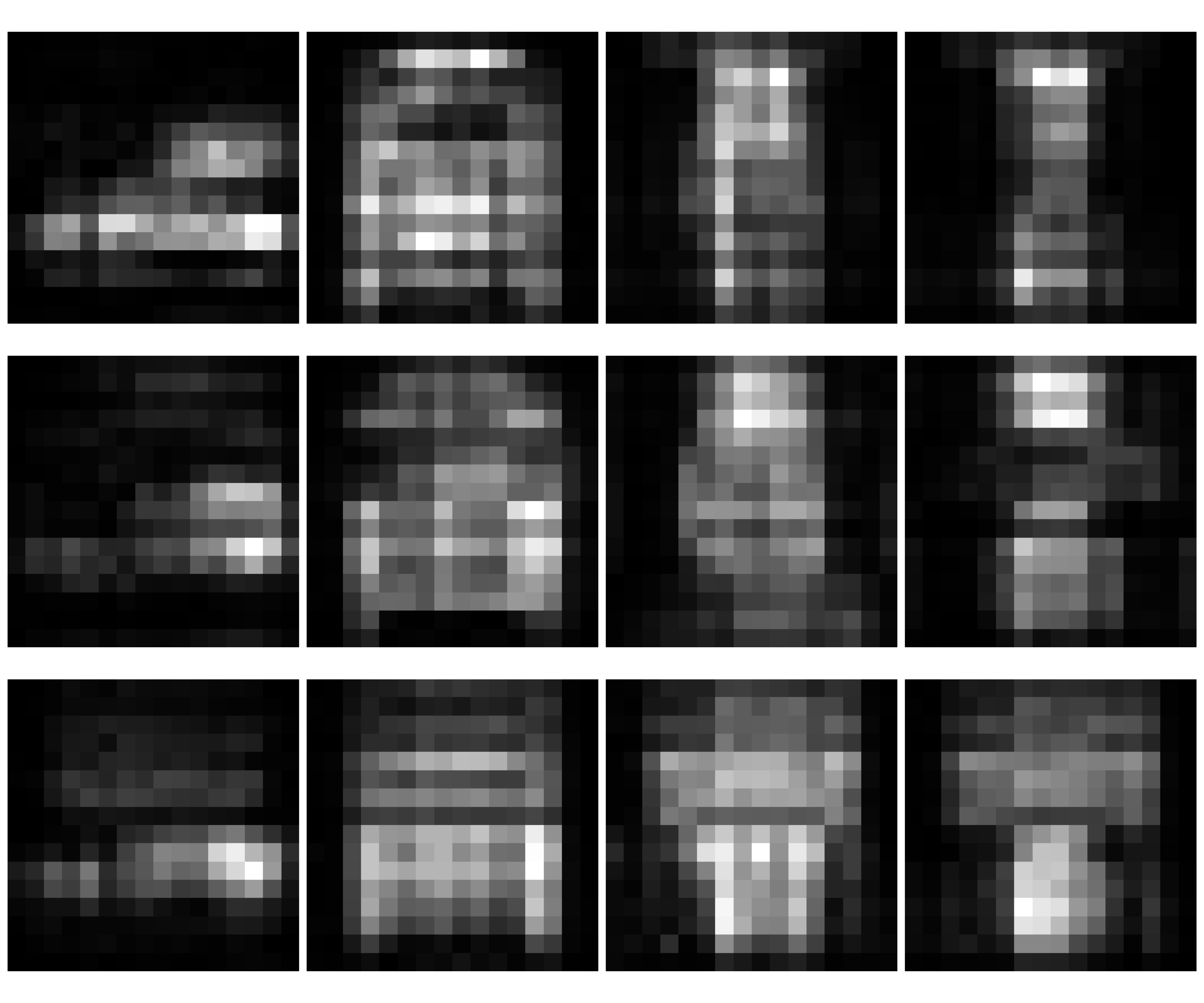}
        \caption{FMNIST}
        \label{fig:fmnistc2}
    \end{subfigure}
    \begin{subfigure}{0.325\linewidth}
        \centering
        \includegraphics[width=\linewidth]{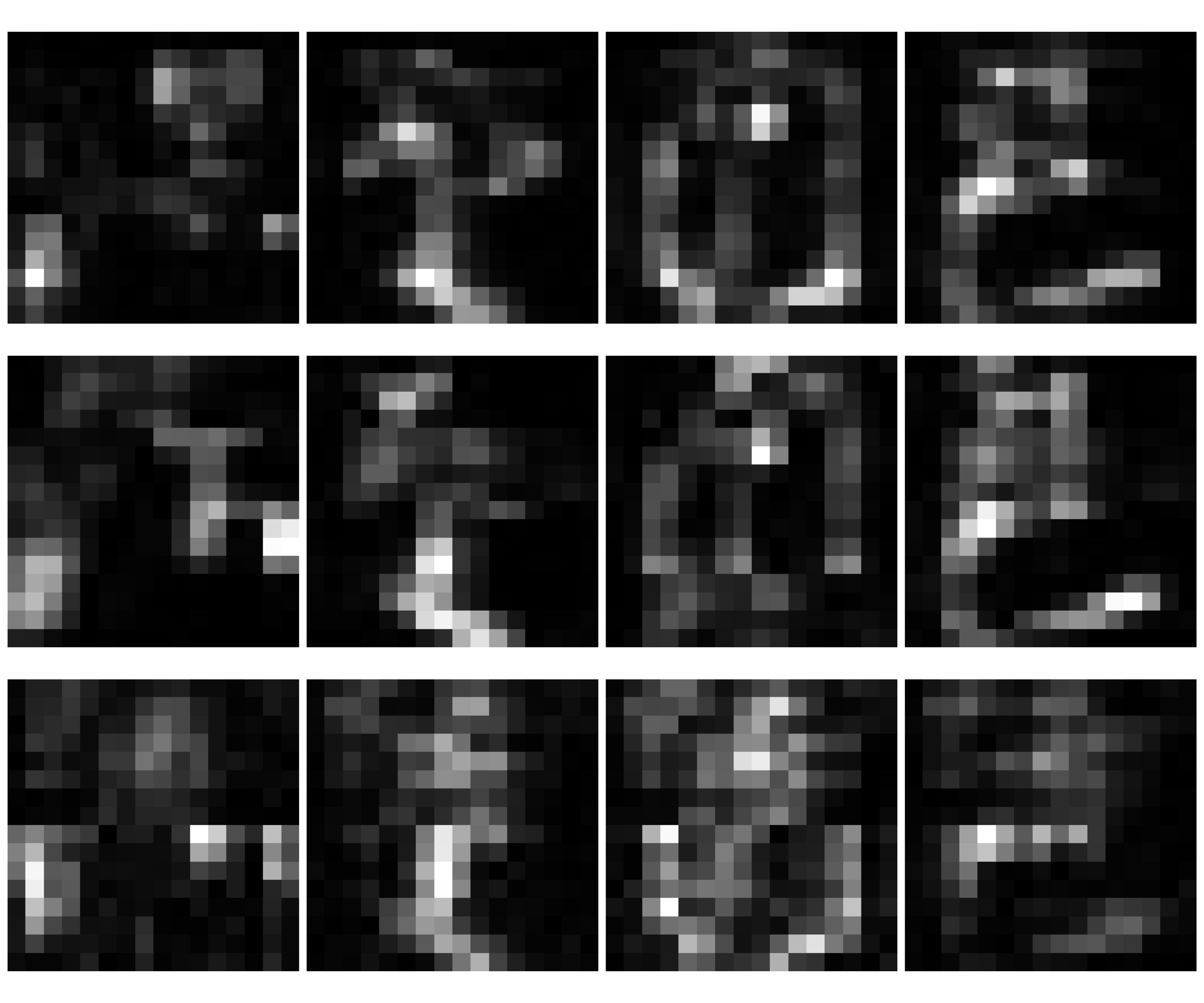}
        \caption{KMNIST}
        \label{fig:kmnistc2}
    \end{subfigure}
    \caption{Results from QMAE at different mask percentages for the (a) MNIST, (b) FMNIST, and (c) KMNIST datasets. Each row showcases the QMAE results at 12.5\% (top row), 25\% (middle row), and 50\% mask (bottom row), respectively.}
    \label{fig:comp}
\end{figure*}

We have also explored the reconstruction quality by varying the mask percentages in the QMAE architecture, shown in Figure~\ref{fig:comp}. This mask percentage represents a tradeoff between the size of the mask and reconstruction quality. Specifically, the reconstruction results for each dataset with 12.5\%, 25\%, and 50\% masks are shown by the images in the top row, middle row, and bottom row, respectively. As expected, 12.5\% mask provides better reconstruction quality compared to 25\% mask. Increasing the mask percentage to 50\% shows that the model is not able to learn the features in the image and produces noise. Therefore, 25\% mask percentage is the best masking value for each of the three datasets.



\begin{figure}
    \begin{subfigure}{\linewidth}
        \centering
        \includegraphics[width=\linewidth]{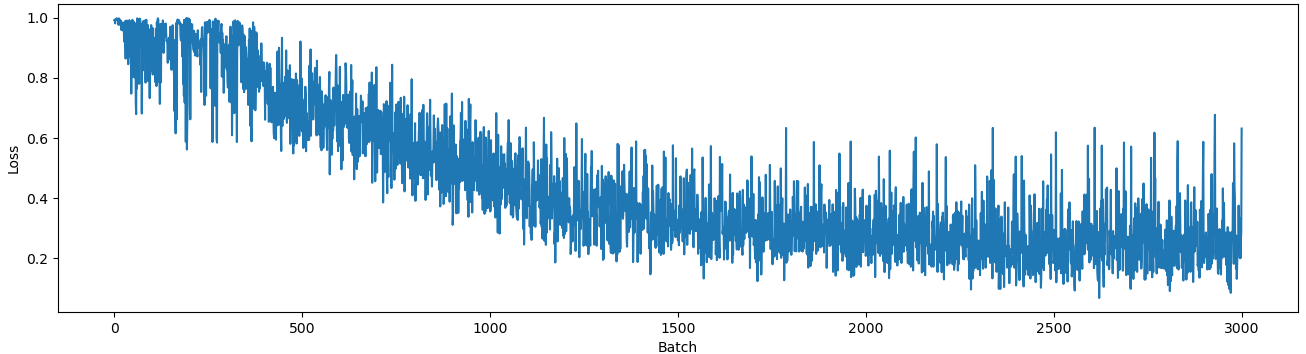}
        \caption{QMAE training loss.} \label{fig:lossqmae}
    \end{subfigure}
    
    \begin{subfigure}{\linewidth}
       \centering
        \includegraphics[width=\linewidth]{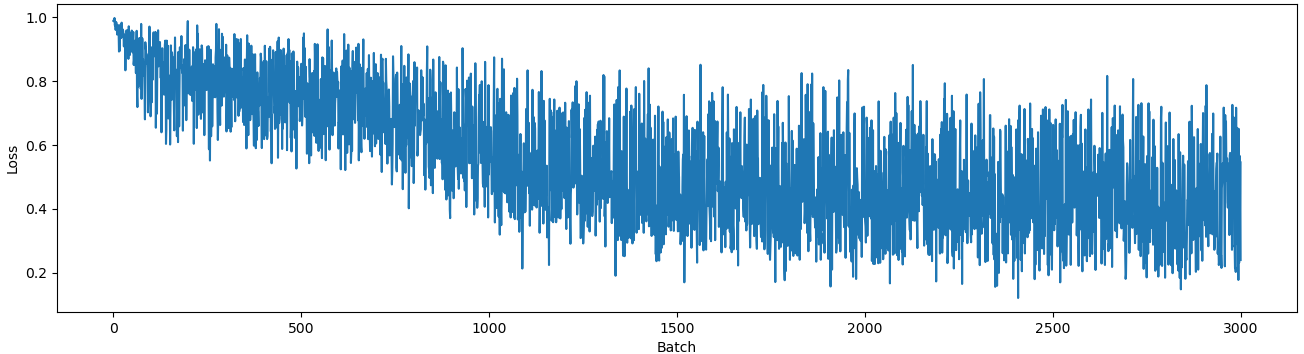}
        \caption{QAE training loss.} \label{fig:lossqae}
    \end{subfigure}
    \caption{Training losses for QMAE (a) and QAE (b). QMAE was able to converge to around 0.25 fidelity loss, whereas QAE struggled to converge and ranged between 0.2 to 0.8 fidelity loss.}
    \label{fig:loss}
\end{figure}

In addition to visual inspection, we quantify the image quality with quantum-based and classical-based metrics of fidelity, cosine similarity (CS), and structured similarity index measure (SSIM). A summary of the obtained metrics for QMAE and QAE under each dataset is presented in Table~\ref{tab:metrics}.

\subsubsection{Quantum-based Metrics}
Like with the loss function, the fidelity between the original image embedded into a circuit and the output from the decoder ansatz can showcase how similar they are. As all data is processed in quantum, this can provide the most accurate similarity metric when the data remains as quantum states. The fidelity is used to train the model through the loss function, as described in Section~\ref{sec:loss}. Figure~\ref{fig:loss} showcases the fidelities for both QMAE and QAE during training. QMAE can converge to 0.25 fidelity loss, while QAE ranges from 0.2 to 0.8. Note that this value is the result of Equation~\ref{eq:loss}. Additionally, the fidelity is measured on 10,000 testing images from each dataset. In the case of MNIST, QMAE has an average fidelity of 0.734 and QAE an average of 0.600. This trend of better average fidelity for QMAE compared to QAE is also observed for FMNIST and KMNIST. For FMNIST, QMAE produced an average fidelity of 0.774 compared to QAE's average of 0.589. QMAE with KMNIST has an average fidelity of 0.670 while QAE has an average of 0.602.

\subsubsection{Classical-based Metrics}
Classical-based metrics include metrics that examine the two targets as images filled with classical pixel values. We utilize both CS and SSIM to calculate the similarity between the original image and the reconstruction resulting from the model. Generally, the images produced by QMAE score higher in similarity with both metrics to the original image compared to the results from QAE. For 10,000 MNIST testing images, QMAE averaged 0.843 CS and 0.446 SSIM, while QAE averaged 0.799 CS and 0.445 SSIM. FMNIST has a similar trend, with QMAE average 0.878 CS and 0.296 CS compared to QAE average 0.771 CS and 0.294 SSIM. While KMNIST QMAE has a better CS of 0.839 compared to QAE with 0.797, QAE outperformed QMAE on SSIM average. The KMNIST QAE was able to reconstruct with a better SSIM average of 0.332 compared to QMAE average of 0.325 SSIM.

\subsection{Comparison of Classification Accuracy}
We have evaluated the quality of the reconstructed images with an image classifier to see if the reconstructed images showcase the features. The image classifier used is a trained ResNet18~\cite{he2016deep} model on $16\times16$ MNIST data. To adapt to the smaller image sizes, the first maxpool layer is replaced in favor of the identity layer. The adjusted ResNet18 model was trained on each dataset individually, resulting in test accuracies of 99.33\% for MNIST, 91.34\% for FMNIST, and 96.07\% for KMNIST. The trained models are used to classify reconstructions from both QMAE and QAE. For each image, both QMAE and QAE were given the same random mask for the best comparison, with the results summarized in Table~\ref{tab:metrics}. On the MNIST test set of 10,000 images, QMAE scored an accuracy of 65.06\%, while QAE scored only 52.20\%, highlighting the fact that QMAE can produce significantly higher quality images than QAE for this dataset. FMNIST and KMNIST performed considerably worse with both QMAE and QAE, indicating issues with recognizing distinct features in each class correctly for both. For these two datasets, the QAE performed marginally better, with 1.26\% and 3.97\% improvements, respectively. 

\begin{table}[h]
    \centering
    \caption{Summary of metrics from QMAE and QAE on 10,000 test samples from each of the datasets.}
    \label{tab:metrics}
    \begin{tabular}{lccccc}
    \toprule
      \textbf{Dataset} & \textbf{Model}   & \textbf{Fidelity} & \textbf{CS} & \textbf{SSIM} & \textbf{Accuracy} \\
      \midrule
      \multirow{2}*{MNIST} & QMAE & \textbf{0.734} & \textbf{0.843} & \textbf{0.446} & \textbf{65.06\%} \\
       & QAE & 0.600 & 0.799 & 0.445 & 52.20\% \\
       \midrule
       \multirow{2}*{FMNIST} & QMAE & \textbf{0.774} & \textbf{0.878} & \textbf{0.296} & 16.15\%\\
       & QAE & 0.589 & 0.771 & 0.294 & \textbf{17.41\%}\\
       \midrule
       \multirow{2}*{KMNIST} & QMAE & \textbf{0.670} & \textbf{0.839} & 0.325 & 11.47\%\\
       & QAE & 0.602 & 0.797 & \textbf{0.332} & \textbf{15.44\%}\\
       \bottomrule
    \end{tabular}
\end{table}

\section{Conclusion}\label{sec:conc}
This paper proposed quantum masked autoencoders (QMAE), a new quantum machine learning architecture to learn the features in masked out information in the original data sample. On image datasets, QMAE can successfully reconstruct missing information that is coherent with the image that is present. This is evidenced by having greater similarity metrics in both quantum and classical measurements compared to state-of-the-art methods, such as quantum autoencoders. The results also highlighted that QMAE reconstructions can present more features of the original dataset, where a classifier can predict with greater accuracy compared to QAEs, such as with MNIST.

\balance

\bibliographystyle{IEEEtran}
\bibliography{IEEEabrv,refs.bib}

\end{document}